\documentclass[article,a4paper,oneside,10pt]{memoir}

\usepackage{amsmath,amsthm}
\usepackage{amssymb}
\usepackage{amstext}
\usepackage{graphicx}
\usepackage{stmaryrd}

\usepackage{color}

\usepackage{pict2e}

\usepackage{xparse}

\makeatletter
\DeclareRobustCommand{\lintprod}{%
  \mathbin{\mathpalette\int@prod{(0,0)(0.8,0)(0.8,0.6)}}%
}
\DeclareRobustCommand{\rintprod}{%
  \mathbin{\mathpalette\int@prod{(0.1,0.6)(0.1,0)(0.9,0)}}}

\newcommand{\int@prod}[2]{%
  \begingroup
  \sbox\z@{$\m@th#1+$}%
  \setlength\unitlength{\wd\z@}%
  \linethickness{0.09\unitlength}%
  \begin{picture}(1,1)
  \roundcap
  \polyline#2
  \end{picture}%
  \endgroup
}
\makeatother

\usepackage[retainorgcmds]{IEEEtrantools}

\usepackage{tikz}

\newcommand{\Eb}{{\mathbf E}}

\newcommand{\Rb}{{\mathbf R}}

\newcommand{\ebf}{{\mathbf e}}

\newcommand{\fb}{{\mathbf f}}

\newcommand{\xb}{{\mathbf x}}

\newcommand{\vb}{{\mathbf v}}

\DeclareDocumentCommand \spinori { o } {%
  \IfNoValueTF {#1} {%
    \sigma%
  }{%
    \sigma_{#1}%
  }%
}

\DeclareDocumentCommand \indG { o o } {%
  \IfNoValueTF {#1} {%
    \iota%
  }{%
    \IfNoValueTF {#2} {%
      \iota_{#1}%
    }{%
      \iota_{#1}(#2)%
    }
  }%
}

\DeclareDocumentCommand \indGc { o o } {%
  \IfNoValueTF {#1} {%
    \bar\iota%
  }{%
    \IfNoValueTF {#2} {%
      \bar\iota_{#1}%
    }{%
      \bar\iota_{#1}(#2)%
    }
  }%
}

\DeclareDocumentCommand \coefG { o o } {%
  \IfNoValueTF {#1} {%
    \gamma%
  }{%
    \IfNoValueTF {#2} {%
      \gamma_{#1}%
    }{%
      \gamma_{#1}(#2)%
    }
  }%
}

\newcommand{\abf}{\mathbf{a}}
\newcommand{\bbf}{\mathbf{b}}
\newcommand{\Bbf}{\mathbf{B}}
\newcommand{\vbf}{\mathbf{v}}
\newcommand{\wbf}{\mathbf{w}}
\newcommand{\zbf}{\mathbf{z}}

\newcommand{\emset}{\mathbf{T}_\text{em}}

\DeclareDocumentCommand \act { o } {%
  \IfNoValueTF {#1} {%
    \mathcal S%
  }{%
    \mathcal S_{\text{#1}}%
  }%
}

\DeclareDocumentCommand \ld { o } {%
  \IfNoValueTF {#1} {%
    \mathcal L%
  }{%
    \mathcal L_{\text{#1}}%
  }%
}

\newcommand{\drm}{\,\mathrm{d}}

\newcommand{\deltabf}{{\boldsymbol \partial}}

\newcommand{\xbpert}{\pmb{\varepsilon}}

\newcommand{\mf}{{\mathbf F}}

\newcommand{\sd}{{\mathbf J}}

\newcommand{\vp}{{\mathbf A}}

\newcommand{\gf}{{\mathbf G}}

\newcommand{\mfbis}{\mathbf{\bar{F}}}
\newcommand{\sdbis}{{\mathbf{\bar{J}}}}
\newcommand{\vpbis}{{\mathbf{\bar{A}}}}

\newcommand{\Gb}{\mathbf{G}}
\makeatletter
\def\trMs{\@ifnextchar[{\@with}{\@without}}
\def\@with[#1]{\Gb_{\xbpert}^{#1}}
\def\@without{\Gb_{\xbpert}^s}
\makeatother

\DeclareMathOperator{\gr}{gr}

\newcommand{\len}[1]{\lvert#1\rvert}

\newcommand{\hodge}{{\scriptscriptstyle\mathcal{H}}}
\newcommand{\hodgeinv}{{\scriptscriptstyle\mathcal{H}^{-1}}}

\newtheoremstyle{question}
  {\topsep}   
  {\topsep}   
  {\upshape}  
  {0pt}       
  {\itshape}  
  {.}         
  {5pt plus 1pt minus 1pt} 
  {\thmname{#1} \thesection.\thmnumber{\itshape#2}\thmnote{(#3)}} 

\makeatletter
    \def\@endtheorem{\hfill$\P$\endtrivlist\@endpefalse }
\makeatother
\theoremstyle{question}

\makeatletter
\let\save@mathaccent\mathaccent
\newcommand*\if@single[3]{%
  \setbox0\hbox{${\mathaccent"0362{#1}}^H$}%
  \setbox2\hbox{${\mathaccent"0362{\kern0pt#1}}^H$}%
  \ifdim\ht0=\ht2 #3\else #2\fi
  }
\newcommand*\rel@kern[1]{\kern#1\dimexpr\macc@kerna}
\newcommand*\widebar[1]{\@ifnextchar^{{\wide@bar{#1}{0}}}{\wide@bar{#1}{1}}}
\newcommand*\wide@bar[2]{\if@single{#1}{\wide@bar@{#1}{#2}{1}}{\wide@bar@{#1}{#2}{2}}}
\newcommand*\wide@bar@[3]{%
  \begingroup
  \def\mathaccent##1##2{%
    \let\mathaccent\save@mathaccent
    \if#32 \let\macc@nucleus\first@char \fi
    \setbox\z@\hbox{$\macc@style{\macc@nucleus}_{}$}%
    \setbox\tw@\hbox{$\macc@style{\macc@nucleus}{}_{}$}%
    \dimen@\wd\tw@
    \advance\dimen@-\wd\z@
    \divide\dimen@ 3
    \@tempdima\wd\tw@
    \advance\@tempdima-\scriptspace
    \divide\@tempdima 10
    \advance\dimen@-\@tempdima
    \ifdim\dimen@>\z@ \dimen@0pt\fi
    \rel@kern{0.6}\kern-\dimen@
    \if#31
      \overline{\rel@kern{-0.6}\kern\dimen@\macc@nucleus\rel@kern{0.4}\kern\dimen@}%
      \advance\dimen@0.4\dimexpr\macc@kerna
      \let\final@kern#2%
      \ifdim\dimen@<\z@ \let\final@kern1\fi
      \if\final@kern1 \kern-\dimen@\fi
    \else
      \overline{\rel@kern{-0.6}\kern\dimen@#1}%
    \fi
  }%
  \macc@depth\@ne
  \let\math@bgroup\@empty \let\math@egroup\macc@set@skewchar
  \mathsurround\z@ \frozen@everymath{\mathgroup\macc@group\relax}%
  \macc@set@skewchar\relax
  \let\mathaccentV\macc@nested@a
  \if#31
    \macc@nested@a\relax111{#1}%
  \else
    \def\gobble@till@marker##1\endmarker{}%
    \futurelet\first@char\gobble@till@marker#1\endmarker
    \ifcat\noexpand\first@char A\else
      \def\first@char{}%
    \fi
    \macc@nested@a\relax111{\first@char}%
  \fi
  \endgroup
}
\makeatother

\newcounter{aside}
   {\refstepcounter{aside}
   \begin{adjustwidth}{\parindent}{-1.5\parindent}
	\small \textbf{Aside \thechapter.\theaside~#1}
	}
{\par\end{adjustwidth}}

\counterwithout{section}{chapter}
\settrimmedsize{297mm}{210mm}{*}
\settypeblocksize{664pt}{498.13pt}{*}
\setulmargins{3cm}{*}{*}
\setlrmargins{*}{*}{0.75}
\setmarginnotes{17pt}{51pt}{\onelineskip}
\setheadfoot{\onelineskip}{2\onelineskip}
\setheaderspaces{*}{2\onelineskip}{*}
\checkandfixthelayout

\providecommand{\keywords}[1]
{
  \small	
  \textbf{\textit{Keywords---}} #1
}
\providecommand{\MSC}[1]
{
  \small	
  \textbf{\textit{MSC---}} #1
}

\newtheorem{remark}{Remark}

\usepackage{ulem}
\newcommand{\fieldA}{\abf}
\newcommand{\fieldB}{\bbf}
\newcommand{\fieldAi}{a}

\newcommand{\Mat}{\mathbf{A}}
\newcommand{\Mati}{a}
\newcommand{\fieldApert}{\abf_\varepsilon}

\RequirePackage{latexsym}
\RequirePackage[colorlinks,citecolor=blue,urlcolor=blue,linkcolor=blue]{hyperref}

\begin{document}

\title{An Exterior Algebraic Derivation of the Euler--Lagrange Equations from the Principle of Stationary Action
\thanks{This work has been funded in part by the Spanish Ministry of Science, Innovation and Universities under grants TEC2016-78434-C3-1-R and BES-2017-081360. Published in: \textbf{\textit{Mathematics} 9, 2178 (2021). DOI:} \href{https://doi.org/10.3390/math9182178}{10.3390/math9182178}.}
}

\author{\scshape Ivano Colombaro\thanks{ivano.colombaro@upf.edu}, Josep Font-Segura\thanks{josep.font@ieee.org}, Alfonso Martinez\thanks{alfonso.martinez@ieee.org} \thanks{Authors are with the Department of Information and Communication Technologies, Universitat Pompeu Fabra, Barcelona, Spain.}}

\maketitle

\begin{abstract}
In this paper, we review two related aspects of field theory: the modeling of the fields by means of exterior algebra and calculus, and the derivation of the field dynamics, i.e., the Euler--Lagrange equations, by means of the stationary action principle. In contrast to the usual tensorial derivation of these equations for field theories, that gives separate equations for the field components, two related coordinate-free forms of the Euler--Lagrange equations are derived. These alternative forms of the equations, reminiscent of the formulae of vector calculus, are expressed in terms of vector derivatives of the Lagrangian density. The first form is valid for a generic Lagrangian density that only depends on the first-order derivatives of the field. The second form, expressed in exterior algebra notation, is specific to the case when the Lagrangian density is a function of the exterior and interior derivatives of the multivector field. As an application, a Lagrangian density for generalized electromagnetic multivector fields of arbitrary grade is postulated and shown to have, by taking the vector derivative of the Lagrangian density, the generalized Maxwell equations as Euler--Lagrange equations.
\end{abstract}


\MSC{primary 37J05; secondary 15A75}

\section{Introduction} \label{sec-intro}

In classical mechanics, the action is a scalar quantity, with units of energy $\times$ time, that encodes the dynamical evolution of a given physical system; mathematically, the action is given by an integral functional of the trajectory or dynamical path (or an integral of the Lagrangian density for field theories) followed by the physical system over space-time. The principle of stationary action states that the actual dynamical path followed by the system, subject to some appropriate boundary constraints, possibly at infinity, corresponds to a stationary point of the action \cite{feynman1977lecturesPhysicsvol2} (Ch.\ 19), \cite{landau1982classicalTheoryFields} (Section 8). An application of the principle yields the Euler--Lagrange equations, which describe the dynamics of the system \mbox{\cite{zee2003quantumFieldTheoryNutshell} (Section I.3)}, \cite{maggiore2005modernIntroductionQFT} (Section 3.1), \cite{weinberg1995quantumTheoryFieldsvol1} (Section 7.2).
The historical development of the stationary-action principle---in essence, a far-reaching generalization of Fermat's principle---that states that light follows the shortest-time path between two points is described in detail in \cite{lanczos1970variationalPrinciplesMechanics} (Section X).

This paper revisits the derivation of the Euler--Lagrange equations for field theories from the principle of stationary action from the point of view of exterior algebra and calculus. There exist several alternative mathematical representations for the fields, ranging from the original vector calculus by Gibbs \cite{Gibbs} and Heaviside to geometric and Clifford algebras \cite{Clifford}, where vectors are replaced by multivectors and 
operations such as the cross and the dot products subsumed in the geometric product; a modern perspective on the use of geometric algbra in physics is given in~\cite{doran2003geometricalgebra}. Early in the 20th century, tensors such as the Faraday tensor in electromagnetism were quickly and almost universally adopted as the natural mathematical representation of fields in space-time \cite{Ricci-Levi-Civita} (pp.~135--144). In parallel, mathematicians such as Cartan generalized the fundamental theorems of vector calculus i.e.,~Gauss, Green, and Stokes, by means of differential forms \cite{Cartan}.
Later on, differential forms were used in Hamiltonian mechanics, e.g.,~to calculate trajectories as vector field integrals \cite{arnold1989mathematicalMethods} (pp.~194--198).
Since differential forms may be seen as the circulation or flux over appropriate space-time regions of multivector fields, it may be preferable in some contexts to directly study the multivector fields.
Therefore, we build our analysis on the exterior algebra originally developed by Grassmann \cite{grassmann1862extensionTheory}, which has comparatively received little attention in the literature and leads to simple formulae that merge the simplicity and intuitiveness of standard vector calculus with the power of tensors and differential forms~\mbox{\cite{colombaro2019introductionSpaceTimeExteriorCalculus,colombaro2020generalizedMaxwellEquations}. 
}

In Section \ref{sec:exterior_algebra}, we provide the necessary background on exterior algebra and calculus, including the important notion of multivector-valued derivative with respect to a vector $\vbf$. Then, we obtain in Section \ref{sec:euler_lagrange_equations} two related coordinate-free forms of the Euler--Lagrange equations for the dynamics of a multivector field $\fieldA$ of grade $r$ as vector derivatives of the Lagrangian density $\mathcal{L}$. 
Our work is related to the geometric--algebraic multivectorial formulation of the Euler--Lagrange equations in~\cite{lasenby1993multivector} (Equations~(4.7) and (4.8)).
The first form in~\eqref{eq:eleqs_tensor} is valid for a generic Lagrangian density that only depends on the first-order derivatives of the field, more specifically on the tensor derivative $\deltabf\otimes\fieldA$ in \eqref{eq:tensor-deriv-def}, and is given by
\begin{equation}
\partial_\fieldA \ld = \deltabf \times 
\bigl( \partial_{\deltabf \otimes \fieldA} \ld \bigr), \label{eq:eleqs_tensor_intro}
\end{equation}
as a function of the vector and matrix derivatives $\partial_\fieldA \ld$ and $\partial_{\deltabf \otimes \fieldA} \ld $ in \eqref{eq:vector_der_1} and \eqref{eq:vector_der_2}, respectively.
The $(k+n)$-dimensional differential operator $\deltabf$ is defined in~\eqref{eq:deltabf}; together with the matrix product $\times$ defined in~\eqref{eq:matrix_vector_transp}, the operation in the right-hand side generalizes the concept of the divergence of a field.
The second form~\eqref{eq:eleqs_exterior}, expressed in exterior algebra notation, is specific to the case when the Lagrangian density depends only on exterior (denoted by $\deltabf\wedge$; see~\eqref{eq:ext-deriv-def}) and interior derivatives (denoted by $\deltabf\lintprod$; see~\eqref{eq:int-deriv-def}) of the multivector field, and is given by
\begin{equation}
\partial_\fieldA \ld = (-1)^{r-1}\deltabf \lintprod \bigl( \partial_{\deltabf \wedge \fieldA} \ld \bigr) + (-1)^{r} \deltabf \wedge \bigl( \partial_{\deltabf \lintprod \fieldA} \ld \bigr), \label{eq:eleqs_exterior_intro}
\end{equation}
where $r$ is the grade of the multivector field $\fieldA$. 
A complementary analysis, which shows the invariance of the action to infinitesimal space--time translations in exterior algebra, was conducted in~\cite{martinez2021sem}, where the stress--energy--momentum tensor is evaluated and profusely discussed.
We conclude the paper in Section \ref{sec:application_maxwell_eqs} with an application of our analysis to a Lagrangian density for generalized electromagnetic multivector fields that leads, by directly taking the vector derivative of the Lagrangian density, to the generalized Maxwell equations for multivector fields of grade $r$ \cite{colombaro2020generalizedMaxwellEquations}. We also provide a short discussion, of independent interest, of a dual form of Maxwell equations where the exterior derivative is replaced by the interior derivative in the definition of the field from the potential.

\section{Fundamentals of Exterior Algebra and Calculus: Notation, Definitions, and Operations} 
\label{sec:exterior_algebra}

\subsection{Multivector Fields}

While our space-time has four space-time dimensions in relativistic terms, it will prove convenient to consider a generic flat space-time $\Rb^{k+n}$ with $k$ temporal dimensions and $n$ spatial dimensions, as this generality allows for a more natural description of the underlying algebraic structure of the equations and of their derivations. Points and position in space-time are denoted by $\xb$, with components $x_i$ in the canonical basis $\smash{\{\ebf_i\}_{i=0}^{k+n-1}}$; by convention, the first $k$ indices, i.e.,~$i=0,\dots,k-1$, correspond to time components while the indices $i = k,\dots,k+n-1$ represent space components. We let space and time coordinates have the same units. Although we shall not make use of this fact, space-time vectors transform contravariantly under changes of coordinates.

In exterior algebra, one considers vector spaces whose basis elements $\ebf_I$ are indexed by lists $I = (i_1,\dots,i_m)$ drawn from $\mathcal{I}_m$, the set of all ordered lists with $m$ nonrepeated indices, with $m \in \mathcal{I} = \{0,1\dots,k+n$\}. Later on, in~\eqref{eq:2.4}, we express the basis elements $\ebf_I$ in terms of the vectorial canonical basis $\ebf_i$, for an ordered list $i_1,\dots,i_m$.
These vectors, which we identify with fields, live in the tangent space and transform covariantly under changes of coordinates \cite{lovelock1989tensorsDifferentialForms} (Ch.\ 2), \cite{flanders1989differentialForms} (Ch.\ V).
We refer to elements of these vector field spaces as multivector fields of grade $m$. While multivector fields do not cover all relevant physical models, e.g., spinor fields or the tensor field in general relativity, they do model a number of interesting cases; for instance, a scalar field is represented by multivectors of grade 0, the electric field, the electromagnetic vector potential and source current by multivectors of grade 1, and the electromagnetic field by a multivector of grade 2.
A multivector field $\fieldA(\xb)$ of grade $m$, possibly a function of the position $\xb$, with components $\fieldAi_I(\xb)$ in the canonical basis $\smash{\{\ebf_I\}_{I\in\mathcal{I}_m}}$ can be written as 
\begin{equation}\label{eq:multivector}
\fieldA(\xb) = \sum_{I\in\mathcal{I}_m} \fieldAi_I(\xb)\ebf_I.
\end{equation}

We denote by $\gr(\fieldA)$ the operation that returns the grade of a vector $\fieldA$ and by $\len{I}$ the length of a list $I$.
The dimension of the vector space of all grade $m$ multivectors is $\smash{\binom{k+n}{m}}$, the number of lists in $\mathcal{I}_m$.

\subsection{Operations on Index Lists}
\label{sec:list_operations}

As the basis elements of multivector fields are indexed by lists $I$, it proves convenient to define some basic operations on such lists: permutations and their signatures, concatenations (mergers), and subtractions of lists. 

First of all, if the list $I$ is not ordered, let $\sigma(I)$ denote the signature of the permutation sorting the elements of $I$ in increasing order. If the permutation is even (resp.\ odd), the signature is $+1$ (resp.\ $-1$). If the list $I$ contains repeated indices, its signature is $0$.

More generally, for two index lists $I$ and $J$ with respective lenghts $m = \len{I}$ and $m' = \len{J}$, let $(I,J) = \{i_1,\dots,i_m,j_1,\dots,j_{m'}\}$ be the concatenation of the two index lists $I$ and $J$. We let $\sigma(I,J)$ denote the signature of the permutation sorting the concatenated list of $\len{I}+\len{J}$ indices, and let $I+J$, or $\varepsilon(I,J)$ if the notation $I+J$ is ambiguous in a given context, denote the sorted concatenated list, which we refer to as merged list. 

In general, we view the lists as ordered sets, and apply standard operations on sets to the lists. For instance, $I$ is contained in $J$, the list $J\setminus I$ is the result of removing from $J$ all the elements in $I$, while keeping the order. As another example, we denote by $I^c$ the complement of $I$, namely the ordered sequence of indices not included in $I$.
We denote the empty list by $\emptyset$; it holds that $\sigma(\emptyset,K) = \sigma(K,\emptyset) = 1$ for an ordered list $K$, and that $\ebf_{\emptyset} = 1$. 

\subsection{Operations on Multivectors}
\label{sec:multivector_operations}

We next define several operations acting on multivectors; our presentation loosely follows \cite{colombaro2019introductionSpaceTimeExteriorCalculus} (Sections 2 and~3) and \cite{colombaro2020generalizedMaxwellEquations} (Section 2) and is close in spirit and form to vector calculus. Introductions to exterior algebra from the perspective and language of differential forms can be found in \cite{lovelock1989tensorsDifferentialForms,flanders1989differentialForms}. A geometric algebra perspective can be found in~\cite{doran2003geometricalgebra}. 
With no real loss of generality, we define the operations only for the canonical basis vectors, the operation acting on general multivectors being a mere extension by linearity of the former. 

First, the dot product $\cdot$ of two arbitrary grade $m$ basis vectors $\ebf_I$ and $\ebf_J$ is defined as
\begin{equation}\label{eq:dot_multi}
	\ebf_I\cdot\ebf_J = \Delta_{IJ} = \Delta_{i_1 j_1}\Delta_{i_2 j_2}\dots\Delta_{i_m j_m},
\end{equation}	
where $I$ and $J$ are the ordered lists $I = (i_1,i_2,\dots,i_m)$ and $J = (j_1,j_2,\dots,j_m)$ and $\Delta_{ij} = 0$ if $i\neq j$, and we let time unit vectors $\ebf_{i}$ have negative metric $\Delta_{ii} = -1$ and space unit vectors $\ebf_{i}$ have positive metric $\Delta_{ii} = +1$. When $m = 0$, we interpret the dot product in \eqref{eq:dot_multi} as 1 since $\ebf_{\emptyset} = 1$. 

The following operations to be defined are the interior and exterior products, which subsume and generalize the operations of gradient, curl, and divergence of vector calculus to multivector fields. These operations transform pairs of multivectors into a multivector of a different grade, introducing in the process some signs, i.e., $\pm 1$. When these signs are related to the dot product in \eqref{eq:dot_multi}, we explicitly write the signs as quantities such as $\Delta_{IJ}$. Other sign contributions arise from the signatures of permutations ordering lists of indices. A common practice in the literature to deal with these signatures is to write factors such as $(-1)^{\len{I}+\len{J}}$. However, it seems more convenient to explicitly keep track of the lists and write the permutation associated to this factor, e.g., $\sigma(I,J)$, as clearer connections between different formulae can be established by harnessing the power of group theory for permutations.

Let two basis vectors $\ebf_I$ and $\ebf_J$ have grades $m = \len{I}$ and $m' = \len{J}$. As defined in \mbox{Section \ref{sec:list_operations}}, let $(I,J) = \{i_1,\dots,i_m,j_1,\dots,j_{m'}\}$ be the concatenation of the two index lists $I$ and $J$, let $\sigma(I,J)$ denote the signature of the permutation sorting the elements of this concatenated list. 
Then, the exterior product of $\ebf_I$ and $\ebf_J$ is defined as
\begin{equation} \label{eq:ext-prod-def}
	\ebf_I\wedge\ebf_J = \sigma(I,J)\ebf_{I+J}.
\end{equation}

The exterior product is thus either zero or a multivector of grade $\len{I}+\len{J}$, since $\sigma(I,J)=0$ when the lists $I$ and $J$ have elements in common. The unit scalar (multivector of grade 0) is an identity of the exterior product, as $1 \wedge \ebf_I = \ebf_I \wedge 1 = \ebf_I$. 
The exterior product provides a construction of the basis vector $\ebf_I$, with $I$ an ordered list $I = (i_1,\dots,i_m)$, from the canonical basis vectors $\ebf_i$, namely
\begin{equation}\label{eq:2.4}
	\ebf_I = \ebf_{i_1}\wedge\ebf_{i_2}\wedge\dots\wedge\ebf_{i_m}.
\end{equation}

When $I = \emptyset$, we adopt the usual convention that the right-hand side is $1$.

We next define two generalizations of the dot product, the left and right interior products. 
Let $\ebf_I$ and $\ebf_J$ be two basis vectors of respective grades $\len{I}$ and $\len{J}$. The left interior product, denoted by $\lintprod$, is defined as
\begin{equation}	\label{eq:left-int-prod}
	\ebf_I \lintprod \ebf_J = 
	 \begin{cases}  \Delta_{II}\sigma(J\setminus I,I)\ebf_{J\setminus I}, & \text{if } I \subseteq J, \\
	0, & \text{otherwise}.
	 \end{cases}
\end{equation}

Although we might have overloaded the meaning of $\sigma(J\setminus I,I)$ to be zero when $I \nsubseteq J$, we prefer to list the separate cases in \eqref{eq:left-int-prod}.
The vector $\ebf_{J\setminus I}$ has grade $\smash{\len{J}-\len{I}}$ and is indexed by the elements of $J$ not in common with $I$.
The use of the word left represents the fact that $\ebf_I$ acts from the left on $\ebf_J$ and removes the elements in $I$ from $J$.

Analogously, the right interior product, denoted by $\rintprod$, of two basis vectors $\ebf_I$ and $\ebf_J$ is defined as 
\begin{equation} 	\label{eq:right-int-prod}
	\ebf_J \rintprod \ebf_I = 
	\begin{cases}
	\Delta_{II}\sigma(I,J\setminus I)\ebf_{J\setminus I}, & \text{if } I \subseteq J, \\ 
	0, & \text{otherwise}.
	\end{cases}
\end{equation}

As in the previous case, the use of the word right represents the fact that $\ebf_I$ acts from the right on $\ebf_J$ and removes the elements in $I$ from $J$.
The unit scalar (multivector of grade 0) acting from the left (resp.\ right) is an identity of the left (resp.\ right) interior product, as $1 \lintprod \ebf_I = \ebf_I \rintprod 1 = \ebf_I$. 

It proves instructive to evaluate the left and right interior products between two multivectors of the same grade, i.e.,~if $\len{I}=\len{J}$. From \eqref{eq:left-int-prod} and \eqref{eq:right-int-prod}, and taking into account that $\sigma(\emptyset,K) = \sigma(K,\emptyset) = 1$ for an ordered list $K$, and that $\ebf_{\emptyset} = 1$, we see that
\begin{equation}
	\ebf_I \lintprod \ebf_J = \ebf_J \rintprod \ebf_I = \ebf_I \cdot \ebf_J, \qquad \text{if } \len{I}=\len{J},
\end{equation}
supporting the idea that the interior products generalize the dot product. Both interior products are grade-lowering operations, as the interior product is either zero or a multivector of grade $\smash{\len{J}-\len{I}}$. 

Finally, we define the complement of a multivector. For a multivector $\ebf_I$ with grade $m$, its Grassmann or Hodge complement, denoted by $\ebf_I^\hodge$, is the unit $(k+n-m)$-vector 
\begin{equation} \label{eq:hodge-transf}
	\ebf_I^\hodge = \Delta_{II}\sigma(I,I^c)\ebf_{I^c},
\end{equation}
where $I^c$ is the complement of the list $I$ 
and $\sigma(I,I^c)$ is the signature of the permutation sorting the elements of the concatenated list $(I,I^c)$ containing all space-time indices.
In other words, $\ebf_{I^c}$ is the basis multivector of grade $k+n-m$ whose indices are in the complement of $I$. In addition, we define the inverse complement transformation as
\begin{equation} \label{eq:hodge-inv-transf}
	\ebf_I^\hodgeinv = \Delta_{I^c I^c}\sigma(I^c,I)\ebf_{I^c} .
\end{equation}

The interior products are not independent operations from the exterior product, as they can be expressed in terms of the latter, the Hodge complement and its inverse:
\begin{eqnarray}\label{eq:left-int-equiv}
\ebf_I \lintprod \ebf_J = \bigl( \ebf_I \wedge \ebf_J^{\hodge} \bigr)^{\hodgeinv}, \\
\label{eq:right-int-equiv}
\ebf_J \rintprod \ebf_I = \bigl( \ebf_J^{\hodgeinv} \wedge \ebf_I \bigr)^{\hodge}.
\end{eqnarray}

The vector calculus cross product between two vectors in $\mathbf{R}^3$ can be expressed in several alternative ways in terms of the interior, and exterior products and Hodge dual \cite{colombaro2019introductionSpaceTimeExteriorCalculus} (Equation (18)). This fact allows us to distinguish various roles that the cross product takes in Maxwell equations and lies at the origin of generalized electromagnetism described by multivectors in generic flat space-time \cite{colombaro2020generalizedMaxwellEquations}.

\subsection{Matrix Vector Spaces}

We do not need to consider general tensor fields but rather the matrix field (vector) space whose basis elements can be represented as $\wbf_{I_1,I_2}$, where both $I_1$ and $I_2$ are ordered lists of nonrepeated $\ell_1$ and $\ell_2$ elements, respectively. We may identify these basis elements with the tensor product of two multivectors of grade $\ell_1$ and $\ell_2$, namely
\begin{equation}
	\wbf_{I_1,I_2} = \ebf_{I_1}\otimes\ebf_{I_2}.
\end{equation}

The dimension of the vector space spanned by these basis elements is $\binom{k+n}{\ell_1}$$\binom{k+n}{\ell_2}$; the elements of this vector space can be identified with matrices $\Mat$ whose rows and columns are indexed by lists, $I_1 \in \mathcal{I}_{\ell_1}$ and $I_2 \in \mathcal{I}_{\ell_2}$, respectively, 
\begin{equation}
	\Mat = \sum_{I_1\in\mathcal{I}_{\ell_1},I_2\in\mathcal{I}_{\ell_2}}\Mati_{I_1I_2}\wbf_{I_1,I_2}. 
\end{equation}

The transpose of a matrix element $\wbf_{I_1,I_2}$, denoted as $\wbf_{I_1,I_2}^T$, is defined as $\wbf_{I_2,I_1}$. These matrices, the underlying vector space, and the operations that we describe next are fundamental in the study of changes of coordinates in space-time. However, consideration of these changes is beyond the scope of this paper. To any extent, this short section provides a perspective on matrices from the point of view of exterior algebra, highlighting the connections between multivectors and matrices, and bypassing the standard introduction of tensor fields.

As we did with multivectors, we consider the dot product $\cdot$ of two arbitrary matrix basis elements $\wbf_{I_1,I_2}$ and $\wbf_{J_1,J_2}$. This dot product is written
\begin{equation}\label{eq:dot_tensor_general}
	\wbf_{I_1,I_2}\cdot\wbf_{J_1,J_2} = \Delta_{I_1J_1}\Delta_{I_2J_2}.
\end{equation}	

The ordering within the pairs $(I_1,I_2)$ and $(J_1,J_2)$ is important in \eqref{eq:dot_tensor_general}. This dot product, when applied to two matrices, is seen to give their Frobenius inner product, or equivalently, the square of the Frobenius norm (also known as the Hilbert--Schmidt norm) \cite{horn2013matrixAnalysis} when the product is of a matrix with itself.
 
We also define the matrix product $\times$ between two matrix basis elements $\wbf_{I,J}$ and $\wbf_{K,L}$ as
\begin{equation}\label{eq:matrix_vector}
	\wbf_{I,J} \times \wbf_{K,L} =  \wbf_{I,L} \Delta_{JK},
\end{equation}
an operation that coincides with the standard product of two matrices for matrices labeled by spatial indices. 
For square matrices $\Mat$ indexed by grade $m$ multivectors, it is natural to define the matrix inverse (whenever the inverse exists), denoted as $\Mat^{-1}$, such that $\Mat^{-1}\times\Mat = \mathbf{I}_m = \Mat\times\Mat^{-1}$,  where the grade $\ell$ square identity matrix, denoted by $\mathbf{I}_\ell$, is given by
\begin{equation}\label{eq:identity-matrix-m}
	\mathbf{I}_\ell = \sum_{I\in\mathcal{I}_\ell}\Delta_{II}\wbf_{I,I}. 
\end{equation}

Last, we define the matrix product $\times$ between a matrix $\wbf_{I,J}$ and a multivector $\ebf_K$ (or between a multivector $\ebf_K$ and the matrix $\wbf_{J,I}$, i.e., the transpose of $\wbf_{I,J}$) as
\begin{equation}\label{eq:matrix_vector_transp}
	\wbf_{I,J} \times \ebf_K =  \ebf_K \times \wbf_{J,I} =  \ebf_{I} \Delta_{JK},
\end{equation}
a generalization of the idea of multiplication of a row (or column) vector by a matrix.

\subsection{Exterior and Matrix Calculus}
\label{sec:exterior_calculus}

In vector calculus, extensive use is made of the partial time derivative, $\partial_t$, and the nabla operator $\nabla$ of partial space derivatives. In our case, we need the generalization to $(k,n)$ space-time to the differential vector operator $\deltabf$, defined as  $(-\partial_0,-\partial_1,\dots,-\partial_{k-1}, \partial_{k},\dots,\partial_{k+n-1})$, that is,
\begin{equation}\label{eq:deltabf}
	\deltabf = \sum_{i\in\mathcal{I}}\Delta_{ii}\ebf_i\partial_i.
\end{equation}	

As was done in \cite{colombaro2019introductionSpaceTimeExteriorCalculus} (Section 3), we define the exterior derivative, $\deltabf\wedge \fieldA$, of a given multivector field $\fieldA$ of grade $m$ as
\begin{equation}\label{eq:ext-deriv-def}
\deltabf\wedge \fieldA = \sum_{i\in\mathcal{I},I\in\mathcal{I}_m:\, i\notin I} \Delta_{ii}\sigma(i,I)\partial_i \fieldAi_I \,\ebf_{i+I}.
\end{equation}

The grade of the exterior derivative of $\fieldA$ is $m+1$, unless $m = k+n$, in which case the exterior derivative is zero.
In addition, we define the interior derivative, $\deltabf\lintprod \fieldA$, of $\fieldA$ as 
\begin{equation}\label{eq:int-deriv-def}
\deltabf\lintprod \fieldA = 
\sum_{i\in\mathcal{I},I\in\mathcal{I}_m:\, i\in I} \sigma(I\setminus i,i) \partial_i \fieldAi_I\, \ebf_{I\setminus i}.
\end{equation}

The grade of the interior derivative of $\fieldA$ is $m-1$, unless $m = 0$, in which case the interior derivative is zero.

The formulae for the exterior and interior derivatives allow us to recover some standard formulae in vector calculus. For a scalar function $\phi$, its gradient is given by its exterior derivative \mbox{$\nabla\phi = \deltabf\wedge\phi$}, while for a vector field $\vb$, its divergence $\nabla\cdot\vb$ is given by its interior derivative $\nabla\cdot\vb = \deltabf\lintprod\vb$. 
Also, for a vector fields $\vb$ in $\Rb^3$, taking into account \cite{colombaro2019introductionSpaceTimeExteriorCalculus} (\mbox{Equation (18)}), the curl can be variously expressed as
	$\nabla\times\vb = (\nabla\wedge\vb)^\hodgeinv = \nabla\lintprod\vb^\hodgeinv = \nabla\lintprod\vb^\hodge$,
thereby generalizing both the cross product and the curl to grade $m$ vector fields in space-time algebras with different dimensions. 
Specific vector calculus formulae such as that for the divergence of a gradient or the curl of the curl of a vector can be seen as instances of general exterior calculus formulae such as \cite{colombaro2019introductionSpaceTimeExteriorCalculus} (Equation (38)) and \cite{colombaro2020generalizedMaxwellEquations} (Equation (35)),
\begin{gather}
\deltabf\lintprod(\fieldA\wedge\fieldB) =  \fieldA (\deltabf\cdot\fieldB) -(\deltabf\cdot\fieldA)\fieldB, \label{eq:leibniz-der1} \\
\deltabf \cdot (\fieldA \lintprod \fieldB) = (\deltabf \wedge \fieldA) \cdot \fieldB + (-1)^{\gr(\fieldA)} (\deltabf \lintprod \fieldB)\cdot \fieldA, \label{eq:leibniz-der2}
\end{gather}
where in \eqref{eq:leibniz-der1}, $\fieldA$ and $\fieldB$ are 1-vectors, while in \eqref{eq:leibniz-der2}, $\fieldA$ and $\fieldB$ are $(s-1)$-vector and $s$-vector, respectively. 

The exterior and interior derivatives satisfy the property $\deltabf\wedge(\deltabf\wedge \fieldA) = 0 = \deltabf\lintprod(\deltabf\lintprod\fieldA)$, for a general twice-differentiable multivector field $\fieldA$. These identities imply the well-known facts that the curl of the gradient and the divergence of the curl are zero.

The circulation $\mathcal{C}(\fieldA,\mathcal{V}^\ell)$ and the flux $\mathcal{F}(\fieldA,\mathcal{V}^\ell)$ of a multivector field $\fieldA$ over an $\ell$-dimensional space-time hypervolume $\mathcal{V}^\ell$ are defined as integrals of interior products of the field with infinitesimal integration volumes:
\begin{align}
	\mathcal{C}(\fieldA,\mathcal{V}^\ell) &= \int_{\mathcal{V}^\ell}\drm^{\ell}\xb\rintprod\fieldA, \label{eq:circ-def} \\
	\mathcal{F}(\fieldA,\mathcal{V}^\ell) &= \int_{\mathcal{V}^\ell}\drm^{\ell}\xb^{\hodgeinv}\lintprod\fieldA. \label{eq:flux-def}
\end{align}

As a specific example for \eqref{eq:flux-def}, the flux of a field over an $(k+n)$-dimensional hypervolume is the volume integral of the field.
For both of these operations, the interior product in the integrand is expressed as a differential form, which allows us to invoke the theory of differential forms to prove a Stokes theorem. This Stokes theorem relates the circulation (resp.\ the flux) of the field over the boundary of some hypervolume to the circulation (resp.\ flux) over the same hypervolume of the exterior (resp.\ interior) derivative of the multivector field.

We also define the tensor derivative of $\fieldA$, $\deltabf\otimes\fieldA$, of a given vector field $\fieldA$ of grade $m$ as
\begin{equation}\label{eq:tensor-deriv-def}
	\deltabf\otimes \fieldA = \sum_{i\in\mathcal{I},I\in\mathcal{I}_m} \Delta_{ii}\partial_i\fieldAi_I\, \wbf_{i,I},
\end{equation}
where $\wbf_{i,I}$ is a matrix vector space basis element. 

To conclude this section, we define a derivative operator with respect to an element of a vector space, e.g., a multivector field or a matrix. 
A relevant example of vector derivative operator is $\deltabf$, where the derivative is taken with respect to the position vector $\xb$.
In general, the vector derivative operator with respect to a multivector field $\fieldA$ of grade $m$ (resp.\ matrix $\Mat$ of dimensions $\ell_1\times\ell_2$) is a multivector field (resp.\ matrix) denoted by $\partial_{\fieldA}$ (resp.\ $\partial_{\Mat}$) \cite{dwyer1948symbolicMatrixDerivatives} and given by
\begin{align}
	\partial_{\fieldA} &= \frac{\partial }{\partial \fieldA} = \sum_{I\in\mathcal{I}_{m}} \Delta_{II} \ebf_I\frac{\partial}{\partial \fieldAi_I} , \label{eq:vector_der_1} \\
	\partial_{\Mat} &= \frac{\partial }{\partial \Mat} = \sum_{I\in\mathcal{I}_{\ell_1},J\in\mathcal{I}_{\ell_2}} \Delta_{II}\Delta_{JJ} \wbf_{I,J}\frac{\partial }{\partial \Mati_{I,J}}. \label{eq:vector_der_2}
\end{align} 

Specifically, we shall later need the exterior vector derivative of a scalar function $g(\xb)$, denoted by $\partial_{\fieldA} \wedge g(\xb)$ or with some abuse of notation simply by $\partial_{\fieldA}\,g(\xb)$, and given by
\begin{eqnarray}
	\partial_{\fieldA} \wedge g(\xb) = \partial_{\fieldA}\,g(\xb) = \sum_{I\in\mathcal{I}_{m}} \Delta_{II} \ebf_I\frac{\partial g(\xb)}{\partial \fieldAi_I},
\end{eqnarray} 
and similarly for the matrix derivative. This exterior vector derivative is thus some form of generalized gradient. We shall need the derivative of a scalar function given by a quadratic form in the field and/or its interior or exterior derivatives. Let $\fieldA$ and $\fieldB$ represent two vectors of the same grade. Evaluation of the vector derivatives is straightforward and coincides with the infinitesimal calculus expressions \cite{dwyer1948symbolicMatrixDerivatives}:
\begin{eqnarray}
	\partial_{\fieldA} \,(\fieldA\cdot\fieldA) = 2\fieldA \label{eq:der_a_aa} \\
	\partial_{\fieldA} \,(\fieldA\cdot\fieldB) = \fieldB. \label{eq:der_a_ab}
\end{eqnarray}

\section{Principle of Stationary Action: Derivation of the Euler--Lagrange Equations}
\label{sec:euler_lagrange_equations}

\subsection{General Case: Lagrangian Dependent on the Tensor Derivative}

As we briefly reviewed in the Introduction, in classical mechanics, one defines the action $\act$ as a scalar quantity, with units of energy $\times$ time, that encodes the dynamical evolution of a physical system. Mathematically, the action $\act$ is an integral functional of the trajectory or path over space-time, or of the Lagrangian density $\ld(\xb)$ for field theories, followed by the physical system. The principle of stationary action states the the path actually followed by the system, e.g., the field dynamics, corresponds to a stationary point of the action \cite{feynman1977lecturesPhysicsvol2} (Ch.\ 19), \cite{landau1982classicalTheoryFields}

(Section 8). 

In general, the application of the principle of stationary action gives the EulerLagrange equations, which describe the dynamics of the system \cite{zee2003quantumFieldTheoryNutshell} (Section I.3), \cite{maggiore2005modernIntroductionQFT} (Section 3.1), \cite{weinberg1995quantumTheoryFieldsvol1} (Section 7.2).
We start by reviewing how to obtain these equations in coordinate-free form with tensorial notation. Differently from the usual approach, that gives the dynamics for the individual components of the field, our coordinate-free derivation directly works with some twice-differentiable multivector field
 $\fieldA$ of grade $s$ and its tensor derivative $\deltabf\otimes\fieldA$ in \eqref{eq:tensor-deriv-def}. 
 
For a given region $\mathcal{R}$ that comprises the physical system under consideration, let the action $\act(\fieldA)$ be given by 
\begin{equation}\label{eq:action-1}
  \act(\fieldA) = \int_\mathcal{R}\! \drm^{k+n}\xb\, \ld(\fieldA, \deltabf \otimes\fieldA).
\end{equation}

We assume that the region $\mathcal{R}$ is large enough to make the physical system closed, and that the fields decay fast enough over $\mathcal{R}$ so that the flux of the fields over the boundary of $\mathcal{R}$ is negligible. 
We note that the Lagrangian density $\ld$ is a real-valued function of the $\smash{\binom{k+n}{s}}$ components of $\fieldA$ and the $(k+n)\smash{\binom{k+n}{s}}$ components of $\deltabf \otimes\fieldA$, and the Lagrangian density does not depend explicitly on the space-time components.

Let the field $\fieldA$ be infinitesimally perturbed by an amount $\fieldApert$, possibly dependent on the space-time coordinates, so that the field is transformed as $\fieldA \rightarrow \fieldA + \fieldApert$ and the tensor derivative is transformed as $\deltabf \otimes\fieldA \rightarrow \deltabf \otimes\fieldA + \deltabf \otimes\fieldApert$. We assume that $\fieldApert$ is twice differentiable.
We can expand the Lagrangian density in a first-order multivariate Taylor series, where the matrix of partial derivatives with respect to the $(k+n+1)\smash{\binom{k+n}{s}}$ variables is a block matrix having along the diagonal the vector derivatives $\partial_\fieldA \ld$ and $\partial_{\deltabf\otimes\fieldA} \ld$ of the density $\ld$ with respect to the field $\fieldA$ and its tensorial derivative $\partial_{\deltabf\otimes\fieldA}$, defined in \eqref{eq:vector_der_1} and \eqref{eq:vector_der_2}, respectively. Neglecting terms of second and higher order in the perturbation $\fieldApert$ and grouping terms in the Taylor series yields 
\begin{equation}\label{eq:action-2}
  \act(\fieldA + \fieldApert) = \int_\mathcal{R}\! \drm^{k+n}\xb\, \Bigl(\ld + (\partial_\fieldA \ld)\cdot \fieldApert + (\partial_{\deltabf\otimes\fieldA} \ld)\cdot (\deltabf\otimes\fieldApert) \Bigr).
\end{equation}

We may thus evaluate the first-order change of action $\delta \act$ as
\begin{equation}
\delta \act = \act(\fieldA + \fieldApert) - \act(\fieldA) \\
= \int_\mathcal{R}\! \drm^{k+n}\xb\, \Bigl((\partial_\fieldA \ld)\cdot \fieldApert + (\partial_{\deltabf\otimes\fieldA} \ld)\cdot (\deltabf\otimes\fieldApert) \Bigr),\label{eq:variations-1}
\end{equation}
always neglecting all the contributions of order $(\fieldApert)^2$ or higher in the action change. 

Next, we note the following Leibniz product rule, an equality between scalar quantities proved in~Appendix \ref{app:leibniz-rule-eq:rel1}, for a multivector field $\abf$ and a matrix field $\Bbf$ with basis $\wbf_{i,I}$, involving the product $\times$ defined in \eqref{eq:matrix_vector_transp},
\begin{equation} \label{eq:rel1}
\deltabf \cdot (\Bbf \times \abf) = (\deltabf \times \Bbf)\cdot\abf +  \Bbf \cdot (\deltabf \otimes \abf) .
\end{equation}

Choosing $\abf = \fieldApert$ and $\displaystyle\Bbf = \partial_{\deltabf \otimes \fieldA} \ld$ in \eqref{eq:rel1}, we can then rewrite Equation \eqref{eq:variations-1} as
\begin{equation}
\label{eq:variations-2}
\delta \act = \int_\mathcal{R}\!\drm^{k+n}\xb\, \Bigl( \partial_\fieldA \ld
-\deltabf \times \bigl( \partial_{\deltabf \otimes \fieldA} \ld \bigr)\Bigr)\cdot \fieldApert \,
+  \int_\mathcal{R}\! \drm^{k+n}\xb \, \deltabf \cdot \Bigl( \bigl( \partial_{\deltabf \otimes \fieldA} \ld \bigr)\times  \fieldApert \Bigr).
\end{equation}

We identify the second integrand with a flux \eqref{eq:flux-def} over an $(k+n)$-dimensional hypervolume $\mathcal{R}$ and use the Stokes theorem \cite{colombaro2019introductionSpaceTimeExteriorCalculus} (Section 3.5) to rewrite the flux of the interior derivative of a vector field as the flux of the field itself across the region boundary $\partial\mathcal{R}$.
The second integral in \eqref{eq:variations-2} then vanishes if we assume that the field $\fieldA$ and its perturbation $\fieldApert$ vanish sufficiently fast at infinity. 
Under this assumption, if the change of action is zero for any perturbation of the field $\fieldApert$, the integrand in the first summand of \eqref{eq:variations-2} must be identically zero. Setting to zero the quantity between parentheses in the integrand yields the coordinate-free form of the Euler--Lagrangian equations, 
\begin{align}
		\partial_{\fieldA} \ld &= \deltabf \times 
( \partial_{\deltabf \otimes \fieldA} \ld), \label{eq:eleqs_tensor_0} \\
\frac{\partial \ld}{\partial \fieldA} &= \deltabf \times 
\left( \frac{\partial \ld}{\partial (\deltabf \otimes \fieldA)} \right). \label{eq:eleqs_tensor}
\end{align}

Both expressions in \eqref{eq:eleqs_tensor_0} and \eqref{eq:eleqs_tensor} are equivalent since they only differ in the notation for the vector derivative. It is also possible to recover a component form of the Euler--Lagrange equations from \eqref{eq:eleqs_tensor} \cite{zee2003quantumFieldTheoryNutshell} (Section I.3), \cite{maggiore2005modernIntroductionQFT} (Section 3.1), \cite{weinberg1995quantumTheoryFieldsvol1} (Section 7.2). Explicitly writing out the definitions in \eqref{eq:vector_der_1} and \eqref{eq:vector_der_2}, we obtain the standard formula for each $I \in\mathcal{I}_m$,
%
\begin{equation}\label{eq:eleqs_components}
 \frac{\partial \ld}{\partial \fieldAi_I} = \sum_{i\in\mathcal{I}} \partial_i \biggl( \frac{\partial \ld}{\partial(\partial_i \fieldAi_I)} \biggr).
\end{equation}

In general, the use of coordinate-free expression as in \eqref{eq:eleqs_tensor} is closer to the common practice of vector calculus and allows us to better identify the algebraic structure of the underlying equations, which gets obscured when the components are used. Moreover, expressions as \eqref{eq:eleqs_tensor} are better suited to generalizations, or more properly, particularizations, to exterior calculus when the Lagrangian depends on the exterior and interior derivatives of the field, rather than the tensor derivative. This case is explored and analyzed in the next subsection.

\subsection{Derivation of the Euler--Lagrange Equations in Exterior Algebraic Form}

For electromagnetism, the Lagrangian density $\ld(\xb)$ is a function of the vector potential $\vp$, the bivector field $\mf$, and the source density vector $\sd$. Expressed in exterior calculus notation, the Lagrangian density is given by
\begin{eqnarray}
	\ld(\xb) &= -\frac{1}{2}\mf\cdot\mf + \sd\cdot\vp.
\end{eqnarray}
\begin{remark}\label{footnote-14}
If the field is represented by an antisymmetric tensor of rank 2, the factor before $\mf\cdot\mf$ becomes $-\frac{1}{4}$ to account for the repeated sum over pairs of indices \cite{zee2003quantumFieldTheoryNutshell} (Section I.5), \cite{maggiore2005modernIntroductionQFT} (Section 3.5), \cite{landau1982classicalTheoryFields} (Section 27).
\end{remark}

The Lagrangian density depends on the field through the potential $\vp$ and its exterior derivative $\mf = \deltabf\wedge\vp$ \cite{colombaro2020generalizedMaxwellEquations} (Section 3). Instead of using \eqref{eq:eleqs_tensor}, which was derived from the assumption that the Lagrangian density depends explicitly only the field and its tensor derivative, it is worth obtaining the Euler--Lagrange equations when the Lagrangian density is a function of a generic multivector field $\fieldA$ of grade $s$, and its exterior and interior derivatives.

As in \eqref{eq:action-1}, for a given region $\mathcal{R}$ that comprises the physical system under consideration, and assumed to 
be large enough to make the physical system closed so that the fields decay fast enough over $\mathcal{R}$ and the flux of the fields over the boundary of $\mathcal{R}$ is arbitrarily small,
the action $\act(\fieldA)$ is given by the integral
\begin{equation}\label{eq:action-1e}
  \act(\fieldA) = \int_\mathcal{R}\! \drm^{k+n}\xb\, \ld(\fieldA, \deltabf \wedge\fieldA, \deltabf \lintprod\fieldA).
\end{equation}

Again, for an infinitesimal perturbation of the field $\fieldApert$, and neglecting all the contributions of order $(\fieldApert)^2$ or higher in the Taylor expansion of the Lagrangian density and the action, the first-order change in action $\delta \act $ is given by
\begin{eqnarray}
\delta \act = \int_\mathcal{R}\! \drm^{k+n}\xb \Bigl( (\partial_\fieldA\ld)\cdot \fieldApert + (\partial_{\deltabf\wedge\fieldA}\ld)\cdot (\deltabf\wedge\fieldApert) +(\partial_{\deltabf\lintprod\fieldA}\ld)\cdot (\deltabf\lintprod\fieldApert) \Bigr).\label{eq:variations-1e}
\end{eqnarray}

From (35) in \cite{colombaro2020generalizedMaxwellEquations}, given a vector $\abf$ and a vector $\bbf$ of grade $\gr(\abf) + 1$, the Leibniz product rule in~\eqref{eq:leibniz-der2} holds.
Choosing $\abf = \fieldApert$ and $\bbf = \partial_{\deltabf \wedge \fieldA}\ld$ (resp.\ $\abf = \partial_{\deltabf \lintprod \fieldA}\ld$ and $\bbf = \fieldApert$) in the second (resp.\ third) summand inside the integral, substituting these values in \eqref{eq:leibniz-der2} and the result back into \eqref{eq:variations-1e}, we obtain
\begin{eqnarray}
\delta \act &= \int_\mathcal{R}\!\drm^{k+n}\xb \Bigl( \partial_\fieldA\ld
+ (-1)^{s+1}\deltabf \lintprod ( \partial_{\deltabf \wedge \fieldA}\ld ) 
\quad - (-1)^{s-1} \deltabf \wedge ( \partial_{\deltabf \lintprod \fieldA}\ld )
\Bigr)\cdot \fieldApert \notag \\
&+  \int_\mathcal{R}\! \drm^{k+n}\xb \, \deltabf \cdot \Bigl( \fieldApert\lintprod (\partial_{\deltabf \wedge \fieldA}\ld ) 
\quad + (-1)^{s} ( \partial_{\deltabf \lintprod \fieldA}\ld )\lintprod\fieldApert\Bigr).
\end{eqnarray}

In the second integrand, a flux over the $(k+n)$-dimensional region $\mathcal{R}$, the Stokes theorem \cite{colombaro2019introductionSpaceTimeExteriorCalculus} (Section 3.5) allows us to rewrite the flux of the interior derivative as the flux across the region boundary $\partial\mathcal{R}$. As both the field $\fieldA$ and its perturbation $\fieldApert$ vanish sufficiently fast at infinity, the first-order change in action is given by
\begin{equation}
\delta \act = \int_\mathcal{R}\!\drm^{k+n}\xb\,\Bigl( \partial_\fieldA\ld
+ (-1)^{s+1}\deltabf \lintprod ( \partial_{\deltabf \wedge \fieldA}\ld ) - (-1)^{s-1} \deltabf \wedge ( \partial_{\deltabf \lintprod \fieldA}\ld )
\Bigr)\cdot \fieldApert.
\end{equation}
and the principle of stationary action, namely that the first-order change in action identically vanishes, leads to the coordinate-free form of the Euler--Lagrange equations, in one of the two equivalent forms:
\begin{align}
 \partial_\fieldA\ld
&= (-1)^{s}\deltabf \lintprod ( \partial_{\deltabf \wedge \fieldA}\ld ) - (-1)^{s} \deltabf \wedge ( \partial_{\deltabf \lintprod \fieldA}\ld ) \label{eq:eleqs_exterior_0} \\
\frac{\partial \ld}{\partial \fieldA} &= (-1)^{s}\deltabf \lintprod \biggl( \frac{\partial \ld}{\partial (\deltabf \wedge \fieldA)} \biggr) - (-1)^{s} \deltabf \wedge \biggl( \frac{\partial \ld}{\partial (\deltabf \lintprod \fieldA)} \biggr). \label{eq:eleqs_exterior}
\end{align}

It might appear that the tensorial and multivectorial expressions in \eqref{eq:eleqs_tensor} and \eqref{eq:eleqs_exterior} differ. If the Lagrangian density depends on the tensor derivative only through the interior and exterior derivatives, we verify in Appendix \ref{app:tensor-exterior} that both expressions are indeed identical and the following identity holds:
\begin{equation}
\label{eq:tensor-exterior-forms}
\deltabf \times \biggl( \frac{\partial \ld}{\partial (\deltabf \otimes \fieldA)} \biggr) = (-1)^{s}\deltabf \lintprod \biggl( \frac{\partial \ld}{\partial (\deltabf \wedge \fieldA)} \biggr) - (-1)^{s} \deltabf \wedge \biggl( \frac{\partial \ld}{\partial (\deltabf \lintprod \fieldA)} \biggr).
\end{equation}

\section{Application to Generalized Electromagnetism: Maxwell Equations}
\label{sec:application_maxwell_eqs} 
 
\subsection{Generalized Maxwell Equations}

As application of the methods derived in the previous section, we study the generalized Maxwell equations \cite{colombaro2020generalizedMaxwellEquations} and their associated fields. For a given natural number $r$, the Maxwell field $\mf(\xb)$ and the generalized source density $\sd(\xb)$ are respectively characterized by multivector fields of grade $r$ and $r-1$ at every point $\xb$ of the flat $(k,n)$-space-time~\cite{colombaro2020generalizedMaxwellEquations} (Section 3). 
The potential field $\vp(\xb)$ is a multivector field of grade $r-1$ such that 
\begin{equation}	\label{eq:field-potential}
	\mf = \deltabf\wedge\vp.
\end{equation}

If we replace the potential $\vp$ by a new field $\vp' = \vp + \bar\vp + \deltabf\wedge\gf$, where $\bar\vp$ is a constant $(r-1)$-vector and $\gf$ is an $(r-2)$-vector gauge field, the homogenous Maxwell Equation~\eqref{eq:maxwell-Gen2} is unchanged~\cite{colombaro2020generalizedMaxwellEquations} (Section 3). For a given Maxwell field, there is therefore some unavoidable (gauge) ambiguity on the value of the vector potential. 

Scalar fields are given by the vector potential by setting $r = 1$ in Minkowski space-time, namely $k = 1$ and $n = 3$. For classical electromagnetism ($r = 2$, $k = 1$, $n =3$), the bivector field is usually expressed as an antisymmetric tensor of rank 2; electrostatics and magnetostatics are recovered for $k = 0$, $n = 3$, by setting $r = 1$ and $r = 2$, respectively. 
The generalized Maxwell equations for arbitrary values of $r$, $k$, and $n$ are the following pair of coupled differential equations:
\begin{eqnarray} 
\deltabf\lintprod \mf = \sd, \label{eq:maxwell-Gen1}\\
\deltabf\wedge \mf = 0. \label{eq:maxwell-Gen2}
\end{eqnarray}

The interior derivative in~\eqref{eq:maxwell-Gen1} and the exterior derivative in~\eqref{eq:maxwell-Gen2} are respectively defined in \eqref{eq:int-deriv-def} and \eqref{eq:ext-deriv-def}. As we stated in Section \ref{sec:exterior_calculus}, the interior derivative lowers the grade by one, while the exterior derivative increases the grade by one; therefore, \mbox{Equation \eqref{eq:maxwell-Gen1}} is an identity of $(r-1)$-vectors while Equation \eqref{eq:maxwell-Gen2} is an identity of $(r+1)$-vectors.

\subsection{Lagrangian Density for Generalized Electromagnetism}

For electromagnetism, the Lagrangian density $\ld(\xb)$ is a function of the potential $\vp$, the Maxwell field $\mf$, and the source density $\sd$. Expressed in exterior calculus notation, we postulate the generalized Lagrangian density to be
\begin{equation}\label{eq:lagrangian_em}
	\ld(\xb) = \frac{(-1)^{r-1}}{2}\mf\cdot\mf + \sd\cdot\vp.
\end{equation}

For classical electromagnetism ($r = 2$, $k = 1$, $n =3$), if the field is expressed as an antisymmetric tensor of rank 2 the factor before $\mf\cdot\mf$ becomes $-\frac{1}{4}$, see Remark~\ref{footnote-14}. In contrast, for electrostatics ($r = 1$, $k = 0$, $n =3$), the Lagrangian density is given by $\ld = \frac{1}{2}\Eb\cdot\Eb + \rho \phi$, where $\Eb$ is the electric field, $\rho$ the charge density, and $\phi$ is the opposite in sign of the usual electric potential, that is, $\Eb = \deltabf \wedge \phi = \nabla \phi$  \cite{feynman1977lecturesPhysicsvol2} (Ch.\ 19). 

While the Lagrangian in \eqref{eq:lagrangian_em} leads to the generalized Maxwell Equations \eqref{eq:maxwell-Gen1} and \eqref{eq:maxwell-Gen2}, as we shall see in the following section, it is not the most general Lagrangian associated to electromagnetism. Two terms that can be added to it respectively deal with the hypothetical mass of the photon, that is, the Proca term \cite{maggiore2005modernIntroductionQFT} (p.\ 107), \cite{jackson1999classicalElectrodynamics} (Section 12.8), and a gauge-fixing term that appears in the context of quantization of the electromagnetic field \cite{zee2003quantumFieldTheoryNutshell} (Section II.7), \cite{maggiore2005modernIntroductionQFT} (Section 7.1). This general Lagrangian density for electromagnetism is now given by
\begin{equation}\label{eq:lagrangian_em_total}
	\ld(\xb) = \frac{(-1)^{r-1}}{2}\mf\cdot\mf + \sd\cdot\vp - \frac{1}{2}m^2\vp\cdot\vp + \frac{(-1)^{r-1}}{2\xi}(\deltabf\lintprod\vp)\cdot(\deltabf\lintprod\vp),
\end{equation}
where $m$ is the hypothetical photon mass and $\xi$ is a parameter that determines the so-called $R_\xi$ gauge; for $\xi = 1$, we have the Feynman gauge, and in the limit $\xi \to 0$, we have the Landau gauge.

\subsection{Euler--Lagrange Equations}

For Lagrangian densities such as \eqref{eq:lagrangian_em} or \eqref{eq:lagrangian_em_total}, which are essentially quadratic forms in the field and/or its interior or exterior derivatives, evaluation of the vector derivatives is straightforward, as the derivative has the same form as that obtained in infinitesimal calculus for the derivative of a polynomial \eqref{eq:der_a_aa} and \eqref{eq:der_a_ab}. For the Lagrangian density in \eqref{eq:lagrangian_em}, evaluation of the derivatives in the Euler--Lagrange Equation \eqref{eq:eleqs_exterior} give
\begin{align}
	\partial_\vp \ld &= \sd, \\
	\partial_{\deltabf\wedge\vp} \ld &= (-1)^{r-1}(\deltabf\wedge\vp), \label{eq:4.7}
\end{align}
from which the Euler--Lagrange equations themselves \eqref{eq:eleqs_exterior_0}, with $s = r-1$, can be expressed as
\begin{equation}
\label{eq:4.8}
\sd = \deltabf \lintprod (\deltabf\wedge\vp) = \deltabf \lintprod \mf,
\end{equation}
namely the generalized nonhomogenous Maxwell Equation \eqref{eq:maxwell-Gen1} for arbitrary $r$, $k$, and $n$. The homogeneous Maxwell Equation \eqref{eq:maxwell-Gen2} is also satisfied as a consequence of the definition of $\mf = \deltabf\wedge\vp$. 

The exterior algebraic formulation of the Lagrangian and the Euler--Lagrange equations brings the advantage of allowing for a more direct derivation of the Maxwell equations, since evaluation of the vector derivatives mimicks more closely the steps carried out in usual differential calculus to evaluate the derivatives.

The factor $(-1)^{r-1}$ in the Lagrangian density is needed to compensate for the identical term $(-1)^{r-1}$ that appears in the Euler--Lagrange Equation \eqref{eq:eleqs_exterior}. An alternative way of writing the Lagrangian density, without this sign factor, would involve replacing one of the exterior derivatives $\deltabf\wedge\vp$ by a right exterior derivative $\vp\wedge{\deltabf}$,  
where the partial derivative operator is understood to act from the right on the potential. In this case, the skew commutativity of the wedge product, $\deltabf\wedge\vp = (-1)^{r-1}(\vp\wedge{\deltabf})$ \cite{colombaro2019introductionSpaceTimeExteriorCalculus} (Section 2.2), cancels the sign in the Lagrangian density and results in a somewhat neater expression for it.

As for the Lagrangian density in \eqref{eq:lagrangian_em_total}, evaluation of the derivatives in the Euler--Lagrange Equation \eqref{eq:eleqs_exterior} give
\begin{align}
	&\partial_\vp \ld = \sd -m^2 \vp, \\
	&\partial_{\deltabf\wedge\vp} \ld = (-1)^{r-1}(\deltabf\wedge\vp), \\
	&\partial_{\deltabf\lintprod\vp} \ld = (-1)^{r-1}\frac{1}{\xi}(\deltabf\lintprod\vp),
\end{align}
from which the Euler--Lagrange Equation \eqref{eq:eleqs_exterior_0} with the Proca and quantization $R_\xi$-gauge terms become
\begin{eqnarray}\label{eq:genME-proca-Rxi}
	\deltabf \lintprod (\deltabf\wedge\vp) + m^2\vp = \sd + \frac{1}{\xi}\deltabf\wedge(\deltabf\lintprod\vp).
\end{eqnarray}

Using the relationship (34) in \cite{colombaro2020generalizedMaxwellEquations}, we may rewrite \eqref{eq:genME-proca-Rxi} in an alternative form with a wave equation,
\begin{eqnarray}\label{eq:genME-proca-Rxi-v2}
	(-1)^{r-1}(\deltabf\cdot\deltabf)\vp + m^2\vp = \sd + \biggl(\frac{1}{\xi}-1\biggr)\deltabf\wedge(\deltabf\lintprod\vp),
\end{eqnarray}
which somewhat simplifies in the Feynman gauge, for which $\xi = 1$.

\subsection{Dual Generalized Maxwell Equations}

An interesting dual form of Maxwell equations is obtained by swapping the roles played by the interior and exterior derivatives in the Lagrangian density and the Maxwell equations themselves. 
Let the ``potential'' $\vpbis$ and ``source density'' $\sdbis$ be multivectors of grade $s$, and let us define a dual Maxwell field $\mfbis$ of grade $r = s-1$ by $\mfbis = \deltabf\lintprod \vpbis$. 
The Lagrangian density is now given by
\begin{align}
\ld(\xb) &= \frac{(-1)^{r}}{2}(\deltabf\lintprod\vpbis)\cdot(\deltabf\lintprod\vpbis) + \sdbis\cdot\vpbis \\
  &= \frac{1}{2}(\deltabf\lintprod\vpbis)\cdot(\vpbis\rintprod\deltabf) + \sdbis\cdot\vpbis, \label{eq:65}
\end{align}
where we used the relationship between left and right interior derivatives $\deltabf\lintprod\vpbis = (-1)^{s+1}(\vpbis\rintprod\deltabf)$ to write \eqref{eq:65}. Direct evaluation of the Euler--Lagrange Equation \eqref{eq:eleqs_exterior}, with $r = s-1$, gives
\begin{equation}
\sdbis = \deltabf \wedge \mfbis.
\end{equation}

This nonhomogeneous ``Maxwell'' equation is complemented by a homogeneous equation $\deltabf \lintprod \mfbis = 0$, itself a consequence of the definition of $\mfbis$ as $\mfbis = \deltabf\lintprod \vpbis$.

As it happened with the generalized Maxwell equations, the exterior algebraic formulation of the Lagrangian and the Euler--Lagrange equations allows for a more direct derivation of the dual Maxwell equations. An interesting question, which we do not dwell upon as it lies beyond the scope of this paper, is whether the physics of the dual Maxwell equations is different from the usual Maxwell equations, or simply involves a transformation of the fields, potential, and source density, with no new phenomena. Along this direction, and leaving the details left as an exercise to the reader, it is relatively easy to verify that one obtains a wave equation relating $\vpbis$ and $\sdbis$ in a ``Lorenz gauge'' where $\deltabf\wedge\vpbis = 0$. Solutions to this wave equation have several independent degrees of freedom or polarizations. The number of these polarizations is $\binom{k+n-2}{r-1}$, as for the standard Maxwell Equation \cite{colombaro2020generalizedMaxwellEquations} (Section 4.3); this number can be justified as the number of possible $(r+1)$-vectors where two dimensions, one temporal and one spatial are fixed, and the remaining $r+1-2 = r-1$ indices have to be filled with $k+n-2$ possible values. We also have a ``Lorentz force'' density $\fb = \sdbis \rintprod \mfbis$ such that a conservation law holds for the stress--energy-momentum tensor $\emset$ of the field \cite{colombaro2020generalizedMaxwellEquations} (Appendix A.2), as for the usual Maxwell equations.

\appendix
\renewcommand{\thesection}{\Alph{section}}

\section{}

\subsection{Proof of the Leibniz Product Rule in \eqref{eq:rel1}}
\label{app:leibniz-rule-eq:rel1}


\renewcommand{\vbf}{\ebf}

Let us consider a multivector field $\abf$ of grade $m$ and a matrix field $\Bbf$ with basis $\wbf_{i,I} = \ebf_i\otimes\ebf_I$, where $\len{I} = m$.
Using the definitions of dot and matrix product \eqref{eq:dot_multi} and \eqref{eq:matrix_vector_transp}, the first term of the right-hand side of~\eqref{eq:rel1} is evaluated as 
\begin{align}
(\deltabf \times \Bbf)\cdot \abf &=  \Biggl( \sum_{i,j\in\mathcal{I},J\in\mathcal{I}_m} \Delta_{ii} \partial_ib_{j,J} \,\ebf_i \times (\ebf_j \otimes \vbf_J )\Biggr) \cdot \Biggl(\sum_{I\in\mathcal{I}_m} a_I \,\vbf_I\Biggr) \label{eq:rel1-0}
\\
&=  \Biggl( \sum_{i\in\mathcal{I},J \in\mathcal{I}_m} \partial_i b_{i,J}\, \vbf_J \Biggr) \cdot \Biggl(\sum_{I \in\mathcal{I}_m} a_I\, \vbf_I\Biggr) \label{eq:rel1-1} \\ 
&= \sum_{i\in\mathcal{I},I \in\mathcal{I}_m} \Delta_{II} a_I\partial_i b_{i,I},
\label{eq:rel1-3}
\end{align}
where in \eqref{eq:rel1-0}, we wrote the components of $\deltabf\Bbf$ and $\abf$, in \eqref{eq:rel1-1}, we computed the matrix product and removed the $j$ index, and in \eqref{eq:rel1-3}, we carried out the dot product and removed the $J$ index.

In turn, the second term in the right-hand side of~\eqref{eq:rel1} can similarly be evaluated using the dot and times products in \eqref{eq:dot_tensor_general} and \eqref{eq:matrix_vector_transp} as
\begin{align}
\Bbf \cdot (\deltabf \otimes \abf) &= \Biggl(\sum_{i\in\mathcal{I},I \in\mathcal{I}_m} b_{i,I}\,\ebf_i \otimes \vbf_I \Biggr) \cdot \Biggl( \sum_{j\in\mathcal{I},J \in\mathcal{I}_m} \Delta_{jj}\partial_j a_J \,\ebf_j \otimes \vbf_J\Biggr) \\
&= \sum_{i\in\mathcal{I},I \in\mathcal{I}_m} \Delta_{II} b_{i,I} \partial_i a_I. \label{eq:rel1-2}
\end{align}

Next, using again the definitions of dot and matrix product \eqref{eq:dot_multi} and \eqref{eq:matrix_vector}, the left-hand side of~\eqref{eq:rel1} becomes
\begin{align}
\deltabf \cdot (\Bbf \times \abf) &= 
\Biggl( \sum_{i\in\mathcal{I}} \Delta_{ii}\ebf_i\frac{\partial}{\partial x_i} \Biggr) \cdot 
\Biggl( \sum_{I\in\mathcal{I}_m} \sum_{j\in\mathcal{I},J \in\mathcal{I}_m} a_I b_{j,J} (\ebf_j\otimes\vbf_J) \times \vbf_I \Biggr) 
\\
&=\Biggl( \sum_{i\in\mathcal{I}} \Delta_{ii}\ebf_i\frac{\partial}{\partial x_i} \Biggr) \cdot 
\Biggl( \sum_{j\in\mathcal{I},I \in\mathcal{I}_m} a_I b_{j,I} \Delta_{II}\ebf_j\Biggr) 
\\
&= \sum_{i\in\mathcal{I},I \in\mathcal{I}_m} \Delta_{II} \partial_i\bigl( a_I b_{i,I} \bigr). 
\end{align}

Summing \eqref{eq:rel1-3} and \eqref{eq:rel1-2} and applying the rule for the derivative of a product yields the desired \eqref{eq:rel1}.

\subsection{Identity between Tensorial and Exterior Algebraic Euler--Lagrange Equations}
\label{app:tensor-exterior}

Since both the exterior and interior derivatives are surjective linear functions of the tensor derivative, the respective vector derivatives of the Lagrangian density are related. Each component of the exterior and interior derivatives \eqref{eq:ext-deriv-def} and \eqref{eq:int-deriv-def} is a scalar (affine) function of several distinct components of the tensor derivative. The Lagrangian density depends on the components of the tensor derivative only through these scalar functions. We thus need to compute the derivative of a function $\ld\bigl(g_1(\zbf),\dots,g_\ell(\zbf)\bigr)$, where $\zbf$ stands for a vector with the $\smash{\ell' = (k+n)\binom{k+n}{s}}$ components of the tensor derivative, and $(g_1,\dots,g_\ell)$ are the (differentiable) functions that give the components of the exterior (resp.\ interior) derivative from the tensor derivative, where $\smash{\ell = \binom{k+n}{s+1}}$ (resp.\ $\smash{\ell = \binom{k+n}{s-1}}$). By construction, a given $z_k = \partial_j\fieldAi_J$ appears only in one $g_i(\zbf)$, the $I$ component of either the exterior or the interior derivative. In the former case, $I = j + J$, in the latter case $I = J\setminus j$.
. 

From the definition of partial derivative, and for any $i = 1, \dots, \ell$, we have the relation
\begin{equation}\label{eq:35}
 \frac{\partial \ld}{\partial g_i(\zbf)} = \lim_{h\to 0} \frac{\ld\bigl(g_1(\zbf),\dots,g_i(\zbf)+h,\dots,g_\ell(\zbf)\bigr)-\ld\bigl(g_1(\zbf),\dots,g_\ell(\zbf)\bigr)}{h} .
\end{equation}

Then, assuming that $\frac{\partial g_i(\zbf)}{\partial z_k}\ne 0$ and defining $h_{ik}' = \frac{h}{\partial g_i/\partial z_k}$, we can then write for every value of $k$ such that the partial derivative $z_k = \partial_j\fieldAi_J$ appears $g_i(\zbf)$,
\begin{equation}\label{eq:36}
g_i(\zbf)+h = g_i(\zbf) + h_{ik}' \frac{\partial g_i(\zbf)}{\partial z_k} \simeq g_i(z_1, \dots, z_k + h_{ik}', \dots, z_{\ell'}),
\end{equation}
where we used the differentiablility of the function $g_i$. Substituting \eqref{eq:36} back into \eqref{eq:35} yields
\vspace{-12pt}
\begin{adjustwidth}{-4.6cm}{0cm} \begin{align}
\frac{\partial \ld}{\partial g_i(\zbf)} =
\frac{1}{\partial g_i/\partial z_k}\lim_{h_{ik}'\to 0} \frac{\ld\bigl(g_1(\zbf),\dotsi,g_i(z_1, \dotsi, z_k + h_{ik}', \dotsi, z_{\ell'}),\dotsi,g_\ell(\zbf)\bigr)-\ld\bigl(g_1(\zbf),\dotsi,g_\ell(\zbf)\bigr)}{h_{ik}'} .
\label{eq:37}
\end{align} \end{adjustwidth}

Since the $z_k$ component appears only in one of the functions $g_i$, the limit in \eqref{eq:37} is the partial derivative of the Lagrangian density with respect to the $k$th component of the tensor derivative, for any $k$, that is,
\begin{equation}\label{eq:38}
\frac{\partial \ld}{\partial g_i(\zbf)} =  \frac{1}{\partial g_i/\partial z_k}\frac{\partial \ld}{\partial z_k}.
\end{equation}

We now proceed to evaluate the vector derivative with respect to the exterior derivative. First, we have
\begin{eqnarray}
\frac{\partial \ld}{\partial (\deltabf \wedge \fieldA)} = \sum_{I\in \mathcal{I}_{s+1}} \frac{\partial\ld}{\partial(\deltabf \wedge \fieldA)_I} \Delta_{II} \ebf_{I}, \label{eq:der-lag-wedge-0}
\end{eqnarray}
where the $I$th component of the exterior derivative, $(\deltabf \wedge \fieldA)_I$ is the equivalent of $g_i(\zbf)$ in \eqref{eq:38}. The equivalent of $k$ is any pair of $j$ and $J\in\mathcal{I}_{s}$ such that $I = j +J$ and the corresponding $z_k$ is $\partial_j \fieldAi_J$. The partial derivative $\partial g_i/\partial z_k$ in \eqref{eq:38} is thus $\Delta_{jj}\sigma(j,J)$, of value $\pm 1$,  and we therefore have for any pair of $i$ and $J\in\mathcal{I}_{s}$ such that $I = j +J$ that
\begin{eqnarray}
	\frac{\partial\ld}{\partial(\deltabf \wedge \fieldA)_I} = \frac{1}{\Delta_{jj}\sigma(j,J)}\frac{\partial\ld}{\partial (\partial_j \fieldAi_J)} = \Delta_{jj}\sigma(j,J)\frac{\partial\ld}{\partial (\partial_j \fieldAi_J)}. \label{eq:der-lag-wedge-1}
\end{eqnarray}

Substituting \eqref{eq:der-lag-wedge-1} back in \eqref{eq:der-lag-wedge-0} yields
\begin{eqnarray}
\frac{\partial \ld}{\partial (\deltabf \wedge \fieldA)} = \sum_{I\in \mathcal{I}_{s+1}} \frac{\partial\ld}{\partial (\partial_j \fieldAi_J)} \Delta_{jj}\Delta_{II} \sigma(j,J) \ebf_{I}, \label{eq:der-lag-wedge}
\end{eqnarray}
where $j$ and $J$ are any pair such that $I = j +J$. 
Now, taking the interior derivative of \eqref{eq:der-lag-wedge}, we obtain for any $j, J$ such that $I = j +J$,
\begin{align}
\deltabf \lintprod \biggl(\frac{\partial \ld}{\partial (\deltabf \wedge \fieldA)}\biggr) &= \sum_{i\in\mathcal{I}} \Delta_{ii}\ebf_{i}\partial_i\lintprod\Biggl(\sum_{I\in \mathcal{I}_{s+1}} \frac{\partial\ld}{\partial (\partial_j \fieldAi_J)} \Delta_{jj}\Delta_{II} \sigma(j,J) \ebf_{I}\Biggr)
\label{eq:der-lag-wedge-2} \\
&= \sum_{i\in\mathcal{I},I\in \mathcal{I}_{s+1}: i \in I} \Delta_{ii}\Delta_{II} \partial_i\biggl(\frac{\partial\ld}{\partial (\partial_i \fieldAi_{I\setminus i})}\biggr) \sigma(i,I\setminus i)\sigma(I\setminus i,i)\ebf_{I\setminus i}, \label{eq:der-lag-wedge-3}
\end{align}
where we have selected $j = i$ and, therefore, $J = I \setminus i$. We also note that $\sigma(i,I\setminus i)\sigma(I\setminus i,i) = (-1)^{s}$.

We now evaluate the vector derivative with respect to the interior derivative in an analogous manner,
\begin{eqnarray}
\frac{\partial \ld}{\partial (\deltabf \lintprod \fieldA)} = \sum_{I\in \mathcal{I}_{s-1}} \frac{\partial\ld}{\partial (\partial_j \fieldAi_J)} \Delta_{II} \sigma(J\setminus j,j) \ebf_{I}, \label{eq:der-lag-lintprod}
\end{eqnarray}
where $j$ and $J$ are any pair such that $I = J \setminus j$. 
Now, taking the exterior derivative of \eqref{eq:der-lag-lintprod}, we obtain, for any $j, J$ such that $I = J \setminus j$,
\begin{align}
\deltabf \wedge \biggl(\frac{\partial \ld}{\partial (\deltabf \lintprod \fieldA)} \biggr) &= \sum_{i\in\mathcal{I}} \Delta_{ii}\ebf_{i}\partial_i\wedge \Biggl(\sum_{I\in \mathcal{I}_{s-1}} \frac{\partial\ld}{\partial (\partial_j \fieldAi_J)} \Delta_{II} \sigma(J\setminus j,j) \ebf_{I}\Biggr)
 \label{eq:der-lag-lintprod-2} \\
&= \sum_{i\in\mathcal{I},I\in \mathcal{I}_{s-1}: i \notin I} \Delta_{ii}\Delta_{II} \partial_i \biggl(\frac{\partial\ld}{\partial (\partial_i \fieldAi_{i+I})}\biggr) \sigma(I,i)\sigma(i,I) \ebf_{i+I}, \label{eq:der-lag-lintprod-3}
\end{align}
where we have selected $j = i$ and, therefore, $J = i + I$. We also note that $\sigma(I,i)\sigma(i,I) = (-1)^{s-1}$.

Putting \eqref{eq:der-lag-wedge-3} and \eqref{eq:der-lag-lintprod-3}, as well as the relationships between the product of permutation signatures, back into the right-hand side of \eqref{eq:tensor-exterior-forms} yields the expression
\begin{align}
\sum_{i\in\mathcal{I},I\in \mathcal{I}_{s+1}: i \in I} \Delta_{ii}\Delta_{II}\partial_i \biggl( \frac{\partial\ld}{\partial (\partial_i \fieldAi_{I\setminus i})}\biggr) \ebf_{I\setminus i} + \sum_{i\in\mathcal{I},I\in \mathcal{I}_{s-1}: i \notin I} \Delta_{ii}\Delta_{II}\partial_i \biggl( \frac{\partial\ld}{\partial (\partial_i \fieldAi_{i+I})}\biggr)\ebf_{i+I}.\label{eq:der-lag-4}
\end{align}

Since the basis elements are multivectors with $s$ components, we may rewrite \eqref{eq:der-lag-4} as
\begin{equation}
\sum_{i\in\mathcal{I},I\in \mathcal{I}_{s}: i \notin I} \Delta_{II}\partial_i \biggl( \frac{\partial\ld}{\partial (\partial_i \fieldAi_{I})}\biggr) \ebf_{I} + \sum_{i\in\mathcal{I},I\in \mathcal{I}_{s}: i \in I} \Delta_{II}\partial_i \biggl( \frac{\partial\ld}{\partial (\partial_i \fieldAi_{I})}\biggr)\ebf_{I}.\label{eq:der-lag-5}
\end{equation}

The two summations in \eqref{eq:der-lag-4} might be further combined in a single summation over $I\in\mathcal{I}_{s}$. The resulting expression coincides with the left-hand side of \eqref{eq:tensor-exterior-forms}, which can be expanded using the computation in \eqref{eq:eleqs_components} into
\begin{equation}
\sum_{i,I\in \mathcal{I}_{s}} \Delta_{II} \partial_i \biggl( \frac{\partial \ld}{\partial(\partial_i \fieldAi_I)} \biggr) \ebf_I.
\end{equation}

\bibliographystyle{IEEEtran}	
\bibliography{physics.bib}

\end{document}